\documentclass[prb,twocolumn,superscriptaddress,showpacs,citeautoscript]{revtex4}
\usepackage{graphicx,amsmath}

\begin{document}


\title{Phase space signatures of the Anderson transition}
\author{Andr{\'e} Wobst}
\author{Gert-Ludwig Ingold}
\author{Peter H{\"a}nggi}
\affiliation{Institut f{\"u}r Physik, Universit{\"a}t Augsburg,
Universit{\"a}tsstra{\ss}e 1, D-86135 Augsburg, Germany}
\author{Dietmar Weinmann}
\affiliation{Institut de Physique et Chimie des Mat{\'e}riaux de Strasbourg,
UMR 7504 (CNRS-ULP), 23 rue du Loess, BP 43, F-67034 Strasbourg Cedex 2, France}

\begin{abstract}
We use the inverse participation ratio based on the Husimi function to
perform a phase space analysis of the Anderson model in one, two,
and three dimensions. Important features of the quantum states remain 
observable in phase space in the large system size limit, while they 
would be lost in a real or momentum space description. From perturbative 
approaches in the limits of weak and strong disorder, we find that the 
appearance of a delocalization-localization transition is connected to 
the coupling, by a weak potential, of momentum eigenstates which are 
far apart in momentum space. This is  consistent with recent results 
obtained for the Aubry-Andr{\'e} model and provides a novel view on 
the metal-insulator transition.
\end{abstract}
\maketitle

\section{Introduction}
Phase space concepts are widely used in various areas of physics like
quantum optics \cite{schle01} and quantum chaos \cite{takah85,dittr98} 
while they are rarely employed in condensed matter physics. In this work, 
we use a phase space analysis to address the Anderson metal-insulator 
transition, and demonstrate that such a description is very useful and 
represents a powerful tool to describe and to elucidate how, as a function 
of a parameter, the nature of the eigenstates changes from delocalized to 
localized. 

While delocalized states call for a description in terms of momentum 
eigenstates, in particular in the ballistic regime, real space methods 
are expected to be appropriate in the localized regime. Even though the 
real space wave function in itself contains already the full information 
about a quantum state, a phase space representation may be much better 
suited to display the relevant information, e.g., in the vicinity 
of a delocalization-localization transition, where both, real space and
momentum space features, are expected to play an important role.

The relevance of a phase space description has recently been illustrated 
by comparing the one-dimensional Anderson model and the Aubry-Andr{\'e} 
model. In the first case, already the presence of very weak disorder leads 
to localized states in the thermodynamic limit \cite{mott61}. In the 
quasiperiodic potential of the Aubry-Andr{\'e} model, however, a localization 
transition occurs at a critical potential strength \cite{aubry80}. 
From a phase space analysis, it was concluded that this qualitative
difference between the two one-dimensional models is due to the very 
different couplings of the momentum eigenstates, induced by the disorder 
and the quasiperiodic potential, respectively \cite{ingol02}.

In this work, we study the phase space behavior of the Anderson model 
in one, two and three dimensions and show that the above considerations
are not restricted to one-dimensional models. In
contrary, it turns out that the proposed relation
between the coupling of momentum eigenstates due to a weak potential and
the occurrence of a metal-insulator transition allows to explain why the 
Anderson transition cannot occur in one dimension.

In view of the wealth of known results \cite{krame93}, the Anderson
model is particularly well suited for this kind of study. 
First studies of the one- and two-dimensional 
Anderson model based on the Wehrl entropy \cite{wehrl79,mirba95} had
already demonstrated that the diffusive regime present in two dimensions
becomes apparent in phase space \cite{weinm99}. An extension to the 
three-dimensional Anderson model has become possible by calculating inverse 
participation ratios in phase space instead of entropies \cite{wobst02}. 
Using the phase space analysis, we recover that in the thermodynamic
limit all states are localized in one dimension while two dimensions 
represent the marginal case. In three and higher dimensions, the phase
space behavior provides clear signatures of the Anderson transition where 
states become localized only above a critical disorder strength \cite{abrah79}. 
This allows one to gain a detailed understanding of the phase space
concepts and opens the road towards their application to more 
complicated systems. 

In Sec.~\ref{sec:ipr} we start by introducing the characterization of
quantum states by their inverse participation ratio (IPR) in phase space as 
well as the corresponding quantities in real and momentum space. The Anderson 
model is introduced in Sec.~\ref{sec:model}, and numerical results for its 
phase space behavior in the whole range from the ballistic to the localized 
regime are presented in Sec.~\ref{sec:warb}. The observed features are
discussed in the light of known properties of the eigenstates. Since the 
behavior in the limiting cases of weak and strong disorder turns out to 
depend on the dimensionality and to be indicative of the existence of a
metal-insulator transition, we devote the main part of this paper to a 
detailed investigation of these limits. Perturbative expansions 
for the inverse participation ratios in the different spaces are
presented for  strong disorder in Sec.~\ref{sec:winf} and for the
limit of weak disorder in Sec.~\ref{sec:weak}. Here, a crucial 
dependence of the inverse participation ratio on dimension is 
identified, and related to the structure of the coupling of momentum 
eigenstates by weak disorder. This important property is only 
apparent in phase space while such signatures cannot be extracted 
from the inverse participation ratios neither in real nor momentum space. 
Our interpretation and the relation to the known properties of the 
Anderson model, in particular in the marginal case of two dimensions, 
is confirmed by an analysis of the dependence of the inverse
participation ratio on system size in Sec.~\ref{sec:size}. Finally,
we present our conclusions in Sec.~\ref{sec:concl}. 

\section{Characterization of states}
\label{sec:ipr}

Among the infinite variety of possible phase space representations of 
a quantum state \cite{hille84}, the Husimi \cite{husim40} or 
Q~function \cite{cahil69} is best suited for our purpose because it 
guarantees a positive definite density. This property will allow us to
define an inverse participation ratio in Eq.~(\ref{eq:iprps}) below. 
The positivity is a direct consequence of the definition of the Husimi 
function
\begin{equation}
\rho(\mathbf{x_0}, \mathbf{k_0}) =
  |\langle \mathbf{x_0},\mathbf{k_0}|\psi\rangle|^2 \, ,
\label{eq:husimi}
\end{equation}
where the state $|\psi\rangle$ is projected onto a minimal uncertainty
state $|\mathbf{x_0},\mathbf{k_0}\rangle$ centered around position 
$\mathbf{x_0}$ and momentum $\mathbf{k_0}$ in phase space. The minimal
uncertainty state assumes a Gaussian form both in position and momentum
representation. Its real space wave function reads
\begin{equation}
\langle\mathbf{x}|\mathbf{x_0},\mathbf{k_0}\rangle=
\left(\frac{1}{2\pi\sigma^2}\right)^{d/4}\!\!
\exp\left(-\frac{(\mathbf{x}-\mathbf{x_0})^2}{4\sigma^2}+
          i\mathbf{k_0}\cdot\mathbf{x}\right)\, .
\label{eq:koh}
\end{equation}
In the definition (\ref{eq:husimi}) of the
Husimi function, the width $\sigma$ appearing in (\ref{eq:koh}) determines 
the relative importance of structures in real and momentum space. We adopt 
this definition for lattice models with periodic boundary conditions
provided that $\sigma\ll L$. Here, $L$ is the number of lattice sites in one 
spatial direction and the lattice constant sets the unit length. Throughout 
this paper, we choose $\sigma=\sqrt{L/4\pi}$ which yields an equal width 
of the Gaussian relative to the system size $L$ and the momentum interval 
running from $k=-\pi$ to $\pi$. 

Since we are ultimately interested in the thermodynamic limit, $L\to\infty$,
let us first discuss the dependence on system size of the phase space 
resolution provided by the Husimi function. Since the $d$ spatial components 
are independent of each other, it is sufficient to consider the one-dimensional
case. For our choice of $\sigma$, the Gaussian smearing arising from the 
projection onto a minimal uncertainty state affects areas in phase space 
which contain of the order of $\sqrt{L}\times\sqrt{L}$ grid points. Structures 
appearing on smaller scales cannot be resolved. However, even though the 
absolute resolution degrades, relative to the size of the system the 
resolution becomes increasingly better as the system size is increased. 
This holds for any $\sigma$ which scales with system size like $L^{\alpha}$
where $0<\alpha<1$. In contrast, the limiting cases $\alpha=0$ and $\alpha=1$ 
behave quite differently. For $\alpha=0$, we have optimal resolution in real 
space but cannot resolve phenomena in momentum space, even in the thermodynamic 
limit.  The opposite is true for $\alpha=1$ where one would obtain a pure 
momentum space description. Our choice of $\alpha=1/2$
leads to an ideal balance between these two extreme cases, and allows
to track features which rely on both, real and momentum space, to the
thermodynamic limit.

The Husimi function contains a tremendous amount of information about a 
quantum state. It turns out, however, that relevant information can already 
be extracted by considering the inverse participation ratio (IPR) in phase 
space \cite{wobst02}
\begin{equation}
P=\sum_{\mathbf{x},\mathbf{k}}\frac{1}{L^d}[\rho(\mathbf{x},\mathbf{k})]^2\,,
\label{eq:iprps}
\end{equation}
where the sum runs over all phase space points 
$(\mathbf{x},\mathbf{k})$. 
The normalization in (\ref{eq:iprps}) is chosen in
such a way that $P=1$ corresponds to an optimal localization around one
lattice point. In phase space, this is achieved by a minimal uncertainty
state. A distribution of the Husimi density over a larger volume in
phase space corresponds to lower values of $P$.

Although the IPR in phase space (\ref{eq:iprps}) is defined in terms of the 
Husimi function $\rho(\mathbf{x},\mathbf{k})$, it may be calculated directly 
from the wave function \cite{manfr00,wobst02,sugit02}, by means of a 
straightforward generalization of the one-dimensional expression given in 
Refs.~~\onlinecite{manfr00,wobst02}. Such an approach provides significant
numerical advantages and is crucial for the treatment of
higher-dimensional systems.

The IPR in phase space $P$ should be compared with the IPR in real space
which has frequently been employed to describe quantum states in disordered 
systems \cite{thoul74,schre85,hashi92,mirli00}. Here, the state
$\vert\psi\rangle$ is projected onto a Wannier state $\vert\mathbf{x}\rangle$ 
localized on a single site of the lattice. This allows to define the IPR in
real space as
\begin{equation}
P_x=\sum_{\mathbf{x}}|\langle\psi|\mathbf{x}\rangle|^4\,,
\end{equation}
which corresponds to the limit $\sigma\to0$ of the IPR in phase space.
It is also convenient to introduce the IPR in momentum space as
\begin{equation}
P_k=\sum_{\mathbf{k}}|\langle\psi|\mathbf{k}\rangle|^4\,,
\label{eq:iprxk}
\end{equation}
where the basis of momentum eigenstates $|\mathbf{k}\rangle$ is given by 
$\langle\mathbf{x}|\mathbf{k}\rangle=\exp(i\mathbf{k}\cdot\mathbf{x})/L^{d/2}$. 

As will be seen below, even the combined information from the IPRs in real 
and momentum space is not equivalent to the information provided by the
IPR in phase space. However, it was shown in Ref.~~\onlinecite{varga02} that 
by an appropriate Gaussian smearing of the real and momentum space densities 
one can define marginal distributions which allow to reproduce the 
behavior of the IPR in phase space. Unfortunately, this approach does not 
result in a reduction of the numerical effort as compared to the calculation 
of the IPR in phase space.

\section{The Anderson model}
\label{sec:model}
In the following, we shall present a detailed comparison of the IPRs in
real, momentum and phase space by considering the Anderson model for 
a quantum particle in a disordered potential. Its Hamiltonian
\begin{equation}
H=-t\sum_{<\mathbf{x},\mathbf{x'}>}(|\mathbf{x'}\rangle\langle\mathbf{x}|
    +|\mathbf{x}\rangle\langle\mathbf{x'}|)
  +W\sum_nv_n|\mathbf{x}\rangle\langle\mathbf{x}|
\label{eq:ham}
\end{equation}
is defined on a $d$-dimensional square lattice with $L$ sites in each
direction. The energy scale is set by the hopping matrix elements
$t=1$ between nearest neighbor sites $<\!\mathbf{x},\mathbf{x'}\!>$.
In order to avoid boundary effects we choose periodic boundary conditions 
in each direction so that every site has $2d$ nearest neighbors. The on-site 
energies $v_n$ forming the disordered potential are drawn independently from 
a box distribution on the interval $[-1/2; 1/2]$ and $W$ denotes the disorder 
strength.

The structure of the quantum eigenstates of the Anderson model depends 
on the disorder strength. For vanishing disorder,
the eigenstates are plane waves and thus are localized in momentum
space. In the opposite limit of strong disorder, localization in 
real space takes place. In order to describe the behavior of the
states in the whole parameter region, and in particular the transition
between the limiting regimes, it is very useful to work with phase
space quantities which adequately take into account real space as well 
as momentum space properties at the same time.

\section{Inverse participation ratios for the Anderson model}
\label{sec:warb}

\begin{figure}
\includegraphics[width=\columnwidth]{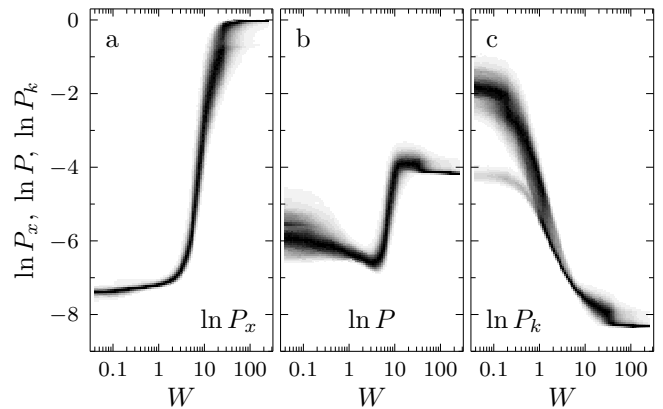}
\caption{Gray scale plot for the distributions of the logarithms of
  the inverse participation ratios in (a) real space, (b) phase space, 
  and (c) momentum space as a function of the disorder strength
  $W$. The data represent $L^2/2$ states around the band 
  center, for 50 different disorder realizations of the two-dimensional 
  Anderson model of size $L=64$.}
\label{fig:ipr2d}
\end{figure}

In order to appreciate the advantage of the phase space approach, 
we start by comparing the IPR in real space, $P_x$, phase space, $P$, and
momentum space, $P_k$, for the two-dimensional Anderson model. In 
Fig.~\ref{fig:ipr2d}, numerical results are shown for a lattice of size 
$64\times 64$. For each given disorder strength $W$, 
we have diagonalized the Hamiltonian (\ref{eq:ham}) for 50 
different disorder realizations $\{v_n\}$, and used $L^2/2$ states
around the band center to calculate distributions of logarithms of 
the IPRs. 

In Fig.~\ref{fig:ipr2d}a we observe a monotonic increase of the real space
IPR with increasing disorder strength $W$. This corresponds to the tendency 
towards localization of the eigenfunctions. According to 
Fig.~\ref{fig:ipr2d}c, the IPR in momentum space simultaneously decreases, 
thereby indicating delocalization in momentum space. This behavior of 
$P_x$ and $P_k$ is an immediate consequence of the system's change 
from the ballistic regime for weak disorder, e.g., localization 
in momentum space, to localized states in real space for strong disorder. 

Since the IPRs in real and momentum space evolve in opposite
directions as a function of the disorder strength, the behavior of the
phase space IPR, which describes the spread of the wave function in
real and momentum space on an equal footing, can be expected to
provide more subtle informations. Indeed, the behavior of the phase
space IPR depends on the details of the model as can be seen by a 
comparison of the one-dimensional Anderson model and the Aubry-Andr{\'e} model
\cite{ingol02}, and within the Anderson model itself, where the
dimensionality plays a crucial role \cite{wobst02}.

For the two-dimensional case, the IPR in phase space 
depicted in Fig.~\ref{fig:ipr2d}b displays a much richer structure
than the IPRs in real and momentum space. In particular, the
dependence on the disorder strength $W$ is non-monotonic, and one finds a 
minimum at an intermediate value of $W$ which can be associated with 
diffusive behavior \cite{wobst02}. 
This non-trivial behavior motivates the following in-depth study of
the Anderson model by means of the IPR in phase space.

\begin{figure}
\includegraphics[width=\columnwidth]{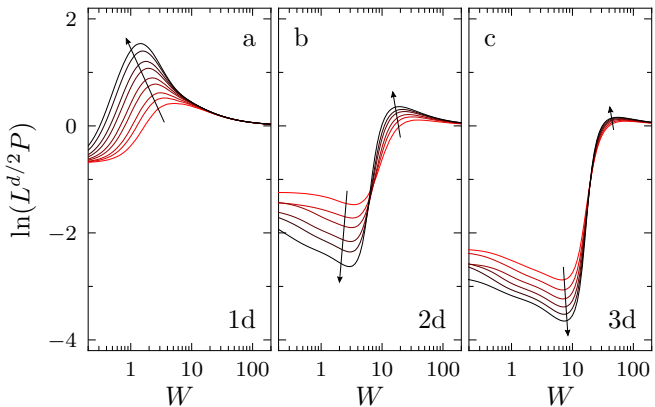}
\caption{Mean IPR in phase space as a function of the disorder strength 
for (a) one, (b) two and (c) three dimensions. The system size $L$
increases from the red to the black curves: (a) $L=128$, 192, 256,
384, 512, 768, 1024, 1536, and 2048; (b) $L=16$, 24, 32, 48, 64, and
96; (c) $L=14$, 16, 18, 20, 22, and 24. The arrows indicate how the 
position $W$ of the extrema shifts with increasing $L$.}
\label{fig:ipr123}
\end{figure}

Figs.~\ref{fig:ipr123}a-c depict the mean 
IPR in phase space for one, two, and three dimensions, respectively, 
for various system sizes $L$. The color changing from red to black 
corresponds to an increasing system size which is also indicated by
the arrows. The data have been scaled with the length dependence
$L^{-d/2}$ of the limiting cases at $W=0$ and $W\to\infty$, cf.\
Sec.~\ref{sec:perturb}. Before giving a detailed discussion of the 
dependence on $L$ in Sec.~\ref{sec:size}, we concentrate on the overall 
behavior as a function of the disorder strength.

One of the most striking aspects of the results presented in
Figs.~\ref{fig:ipr123}a-c is that the behavior of the phase space
IPR at weak disorder depends on the spatial dimension in a crucial way.
While in $d=1$ the IPR increases with increasing $W$, it decreases in 
$d\ge2$. Together with the fact that, independently of the dimension $d$, 
at strong disorder the limiting value for $W\to\infty$ is approached from 
above, this has important consequences for the global behavior 
of the phase space IPR. In $d=1$, the two limits are joined by a peak 
indicating localization in phase space. In contrast, in two and three 
dimensions, $P$ decreases in the regime of small disorder, and assumes a 
minimum indicating a large spreading in phase space
followed by a more or less steep rise towards a maximum 
as can be seen in Figs.~\ref{fig:ipr123}b and c.

The minimum of the phase space IPR in two and higher dimensions can be 
associated with the existence of a diffusive regime where the system size 
is much larger than the mean free path but smaller than the localization 
length. The resulting mixing of the plane waves by the disorder
potential considerably alters the structure of the states and leads to
a spreading both in real and momentum space and thus to a small value
of the phase space IPR. This is reminiscent of the emergence of
quantum chaos and can be confirmed by determining the energy level 
statistics around the minimum of $P$. One indeed finds the
Wigner-Dyson statistics \cite{wobst02} which characterizes the
diffusive (chaotic) regime.

\begin{figure}
\includegraphics[width=\columnwidth]{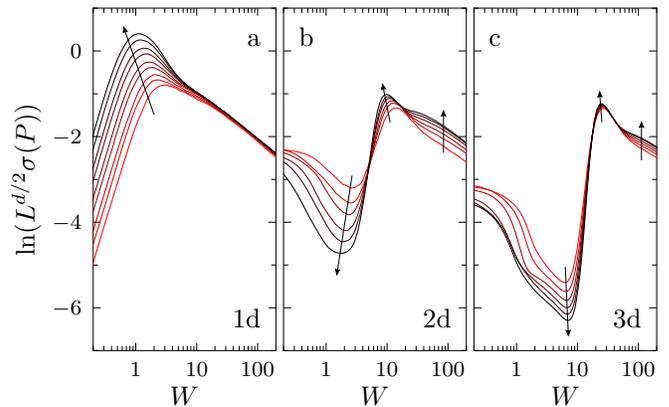}
\caption{The standard deviation $\sigma(P)$ of the IPR in phase space 
as a function of the disorder strength for (a) one, (b) two and (c) 
three dimensions and the same parameters as in 
Fig.~\protect\ref{fig:ipr123}.}
\label{fig:sigma123}
\end{figure}

In addition to the mean value, the distribution of the phase space IPR
at given disorder strength (cf.\ Fig.~\ref{fig:ipr2d}b for the case $d=2$) 
can be characterized by the standard deviation $\sigma(P)$ depicted in
Fig.~\ref{fig:sigma123} for one, two, and three dimensions. 
Here, we have employed the same scaling with system size as 
in Fig.~\ref{fig:ipr123}. 
The overall structure resembles the one found for the mean values. 
The strong suppression of the standard deviation occurring in
the diffusive regime, being particularly pronounced in $d=3$,
indicates that the phase space structure is quite independent 
of the individual states. This confirms once more the universal 
chaotic character of the diffusive states which is expected due to 
the strong mixing present in this regime.

\section{Perturbation theory}
\label{sec:perturb}

The numerical results for the phase space IPR presented in
Figs.~\ref{fig:ipr123} and \ref{fig:sigma123} indicate that the 
changes in the global behavior as a function of the disorder strength 
can be understood in terms of the limiting behavior for strong and, in 
particular, for weak disorder. Therefore, we proceed next to a detailed 
perturbative investigation of the IPRs in these two limits. We start with
the simpler case given by the limit of strong disorder. 
 
\subsection{IPR at strong disorder}
\label{sec:winf}

For $W\to\infty$, all eigenstates are localized on 
single sites in real space. A finite ratio $t/W$ then leads to a coupling to
the nearest neighbor sites due to the kinetic energy in (\ref{eq:ham}).
Such a perturbation can be treated analytically as long as it is sufficient
to take into account only the coupling to the nearest neighbor state
which is closest in energy to the initial site. For the resulting two 
state system, the IPRs may be calculated explicitly. The other nearest 
neighbor sites enter in the calculation only when the disorder average
is performed.

In a first step, we thus focus on two nearest neighbor Wannier states on a
lattice of size $L^d$. The absolute value of the difference between the 
corresponding on-site energies will be denoted by $\Delta$. Then the
effective Hamiltonian for the two level system in the Wannier basis reads
\begin{equation}
H_{\mathrm{TLS}}=
\left(\begin{array}{cc}-\Delta/2&-t\\-t&\Delta/2\end{array}\right)\,.
\label{eq:ham2}
\end{equation}
It is straightforward to determine the two eigenstates and the
corresponding IPRs, which are identical for both states. Introducing the
eigenenergies $\widetilde\Delta=\pm[(\Delta/2)^2+t^2]^{1/2}$, the IPRs
are given by
\begin{equation}
\begin{aligned}
P_x(\Delta)&=1-\frac{t^2}{2\widetilde\Delta^2}\,,\\
P_k(\Delta)&=L^{-d}\left[1+\frac{t^2}{2\widetilde\Delta^2}\right]\,,\\
P(\Delta)&=L^{-d/2}\left[1+\frac{t^2}{2\widetilde\Delta^2}(2\exp(-1/4\sigma^2)
                            -1)\right]\,.
\end{aligned}
\label{eq:iprd}
\end{equation}
In particular, one finds $P_x(0)=1/2$ because for degenerate on-site
potentials the two states are both equally distributed over the two sites.
Furthermore, and consistent with the results of the previous section,
the IPRs in real and momentum space behave oppositely as $t/W$ is
increased. For large system size, the IPR in phase space increases
with $t/W$, just as the IPR in momentum space.

In order to compare with our numerical results, we need to perform a
disorder average. Since the on-site energies are equally distributed
inside the interval $[-W/2; W/2]$, the probability density $p_1$ that 
two neighboring on-site energies differ by $\Delta$ reads 
\begin{equation}
p_1(\Delta)=\frac{2}{W^2}(W-\Delta)\,.
\label{eq:p1}
\end{equation}
The index $1$ indicates that only one nearest neighbor site is
taken into account.

Furthermore, we need to ensure that the energy difference $\Delta$ is the
smallest among the energy differences with all nearest neighbors. Therefore, 
for the remaining $2d-1$ nearest neighbors, the difference in on-site energy 
with respect to the central site should be larger than $\Delta$. The 
probability density for such a $2d$ nearest neighbor configuration is given by
\begin{equation}
\begin{aligned}
p_{2d}(\Delta)&=N^{-1}\,p_1(\Delta)
\left(\int_{\Delta}^W dx\,p_1(x)\right)^{2d-1}\\
&=4d\left(1-\frac{\Delta}{W}\right)^{4d-1}\,,
\end{aligned}
\label{eq:pn}
\end{equation}
where $N$ is a normalization constant.

Within the assumption that we can restrict ourselves to an effective
two level system, we therewith obtain the IPR in phase space
\begin{equation}
P=\int_0^W d\Delta\,p_{2d}(\Delta)P(\Delta)
\end{equation}
together with corresponding expressions for the IPRs in real and momentum 
space.  Making use of (\ref{eq:iprd}) and (\ref{eq:pn}), to leading
order in $t/W$, one obtains
\begin{equation}
\begin{aligned}
P_x&=1-2\pi d\frac{t}{W}\\
P_k&=L^{-d}\left(1+2\pi d\frac{t}{W}\right)\\
P&=L^{-d/2}\left(1+2\pi d\frac{t}{W}[2\exp(-1/4\sigma^2)-1]\right)\, ,
\end{aligned}
\label{eq:winf}
\end{equation}
with corrections of order $(t/W)^2\ln(t/W)$. Configurations where more than
one nearest neighbor site is energetically degenerate with the central site
do not modify the results (\ref{eq:winf}) because the probability to find such 
a configuration vanishes. 

\begin{figure}
\includegraphics[width=\columnwidth]{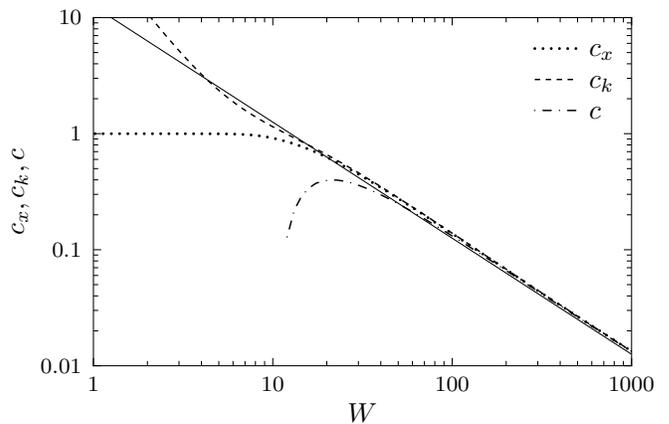}
\caption{Comparison of the perturbative result $2\pi d(t/W)$ (solid
  line) with the numerically computed values for deviations
  of the IPRs from their value at $W=\infty$ in
real space ($c_x$, dotted line), momentum space ($c_k$, dashed line), 
and phase space ($c$, dashed-dotted line), for the parameters of 
Fig.~\ref{fig:ipr2d}.} 
\label{fig:iprwinf}
\end{figure}

In order to compare the numerical data presented in Fig.~\ref{fig:ipr2d}
with the perturbative result, we introduce the quantities
$c_x=1-P_x$, $c_k=L^dP_k-1$, and
$c=(L^{d/2}P-1)/[2\exp(-1/4\sigma^2)-1]$. 
Within the perturbative results of Eq.~\ref{eq:winf}, we have 
$c_x=c_k=c=2\pi d (t/W)$. The numerical results for the
two-dimensional Anderson model are shown in Fig.~\ref{fig:iprwinf}.
The agreement with the leading perturbative results is
remarkably good for disorder strengths down to rather small values of $W$. 
This is particularly true for the IPR in momentum space represented by the
dashed line. The fact that IPRs are by definition positive quantities implies 
that the correction $c_x$ of the IPR in real space depicted by the dotted 
line is limited from above by 1. Therefore, the leading correction to $P_x$ 
given by (\ref{eq:winf}) must fail when $c_x$ reaches this limiting value. 
Finally, the dashed-dotted line corresponding to the phase 
space term $c$ is well described by the leading perturbative
correction according to (\ref{eq:winf}) 
down to $W\approx 30$ for this system of $64\times64$ sites. 

As for the case of two sites, Eq.~(\ref{eq:iprd}), the IPRs in real
and momentum space move in the opposite direction as a function of
$t/W$. Moreover, the IPR in phase space still behaves similarly to the 
momentum space IPR, for the averaged quantities given by
(\ref{eq:winf}). The key to an understanding of this behavior of the 
IPR in phase space lies in the limited resolution provided by the Husimi 
function. Since its spatial resolution is of 
order $L^{1/2}$, changes which occur only on two lattice sites will
not affect the Husimi functions, in particular in the case of large
system sizes. Only the small deviation of the factor 
$2\exp(-1/4\sigma^2)-1$ from 1 can be 
traced back to real space behavior as an incomplete overlap of the Gaussians 
centered at the two sites in question. 
Small scale changes in real space, however, lead to large scale 
changes in momentum space. In the regime discussed above, one observes 
beatings in the momentum space density as a 
consequence of the required orthogonality of the two eigenstates of 
(\ref{eq:ham2}). This effect can be resolved by the Husimi function, so that 
momentum space effects dominate the phase space behavior at strong disorder. 

Finally, the difference of $L^{d/2}$ in the prefactor of the IPRs 
in phase space and momentum space stems from the Gaussian smearing in phase
space which contributes, in our case of spatially well-localized states, a
factor $\sigma\propto L^{1/2}$ for each spatial dimension. 

\subsection{IPR at weak disorder}
\label{sec:weak}

As shown in Fig.~\ref{fig:ipr2d}, the IPRs in real 
and momentum space exchange their qualitative role as compared to the strong 
disorder limit. This is not surprising, because ballistic motion of a quantum 
particle implies the existence of plane waves with well localized momentum 
and delocalization in real space. Exchanging real and momentum space, this 
corresponds to the real space scenario for strong disorder. 

The situation, however, is more complicated in phase space, and the
behavior in the limit of weak disorder, $W\to0$, is by far
more complex. 
Only in one dimension, 
the IPR in phase space can 
indeed be understood in terms of the real space IPR at weak disorder and the 
momentum space IPR at strong disorder.\cite{wobst02} In particular, 
the IPR in phase space increases with increasing disorder strength in
the regime of weak disorder. 
The scenario, however, is very different for two and higher dimensions as 
can already be seen from Fig.~\ref{fig:ipr2d}b where the phase space IPR 
displays a decrease at weak disorder. In this case, the momentum space 
behavior dominates the phase space IPR at both, weak and strong
disorder. Examples of IPRs in real, momentum, and phase space in 
dimensions up to $d=3$ are given in Fig.~4 of
Ref.~~\onlinecite{ingol03}. In the following, we will  
distinguish the cases $d=1$ and $d\ge2$. 

\subsubsection{IPR at weak disorder for $d=1$}

First, we briefly review the phase space properties of the one-dimensional 
Anderson model, which were already discussed in Refs.~~\onlinecite{wobst02} 
and\ \ \onlinecite{ingol02} in some detail. For $W=0$, two plane waves at 
momentum values $k$ and $-k$ are energetically degenerate, and there
is an ambiguity in the choice of the corresponding two basis states. 
We choose symmetric and antisymmetric combinations of the two plane waves,
in order to obtain real wave functions. 
The solutions in the limit $W\to0$ singled out by degenerate 
perturbation theory contain additional phases which, however, do not 
influence any of the discussed IPRs.

In the clean case, $W=0$, one finds $P_k=1/2$ for states with non-vanishing 
momentum. This corresponds to the equally weighted contribution of the two 
momenta $k$ and $-k$. In real space, the sum appearing in the IPR can be 
approximated by an integral and it is sufficient to consider as a 
representative the wave function $\sqrt{2/L}\cos(kx)$. This yields for the 
IPR in real space\footnote{For certain states deviations from this
  result may occur due to the integral approximation, as for example
  in the case $\int_0^Ldx\exp(2\pi ix)=0 \ne \sum_1^L\exp(2\pi ix)=L$.
  These states are of zero measure in the limit of large system size.}
\begin{equation}
\int_0^L\!dx\left(\sqrt{\frac{2}{L}}\cos(kx)\right)^4=\frac{3}{2L}\,.
\label{eq:iprxw01d}
\end{equation}
In phase space the Husimi function resolves the two momenta $k$ and
$-k$ which are well separated for energies around the band center.
While for $W\to\infty$ a single stripe in phase space leads to an
inverse participation ratio $L^{-1/2}$, the two stripes now result in
$L^{-1/2}/2$.

The presence of a disorder potential leads to a coupling of plane waves
with different momenta. In contrast to the opposite case of strong 
disorder, $W\to\infty$, where the coupling of the Wannier states 
occurs only between neighboring sites, for weak disorder $W\to0$, the
coupling of the plane waves is not restricted to neighboring momenta. 
In fact, the averaged matrix element of the disorder potential is 
independent of the momenta of the states involved. 
Within perturbation theory, however,
the energy difference of the states comes into play so that effectively
the coupling to states close in energy (but not necessarily in
momentum) is dominant.

Only for the one-dimensional case, the dispersion relation $E=-2t\cos k$  
implies that real basis states at $W=0$ which are close in energy $E$ are also
close in momentum $k$. As a consequence, only states which are close in 
momentum are efficiently coupled by a weak disorder potential. Although the 
perturbative treatment is more complicated for weak than for strong 
disorder, a qualitative impression of its effect on the phase space 
properties can be obtained in analogy to the case $W\to\infty$ by 
interchanging real and momentum space. Now, because of the limited
resolution of the Husimi function, the coupling to close 
states in momentum space does not have a significant effect while the 
large scale modulation in real space associated with the coupling
affects the Husimi function. Therefore, while the value of the IPR in
phase space for a clean one-dimensional Anderson model at $W=0$ is a 
direct consequence of the localization in momentum space, 
the corrections for finite $W\to 0$ are dominated by real space effects. 

\subsubsection{IPR at weak disorder for $d\ge2$}

Generic ballistic states on a $d$-dimensional cubic lattice display a
$2^dd!$-fold energetic degeneracy. The factor $2^d$ arises from
the degeneracy between momentum vectors 
with different signs of the components, while the factor $d!$ accounts 
for the number of possible permutations of a set of $d$ momentum
values, provided they are all different. For example, in three dimensions all
the 8 different combinations of the signs in 
$\mathbf{k}=(\pm k_1, \pm k_2, \pm k_3)$, and all the 6 permutations
of the momentum components $\{k_1,k_2,k_3\}$ lead to the same energy
$E=-2t(\cos k_1+\cos k_2+\cos k_3)$.
Occasionally, the degeneracies may even be larger. This is the case 
when the same total energy can be achieved by different sets of 
momentum components.

\begin{figure}
\includegraphics[width=\columnwidth]{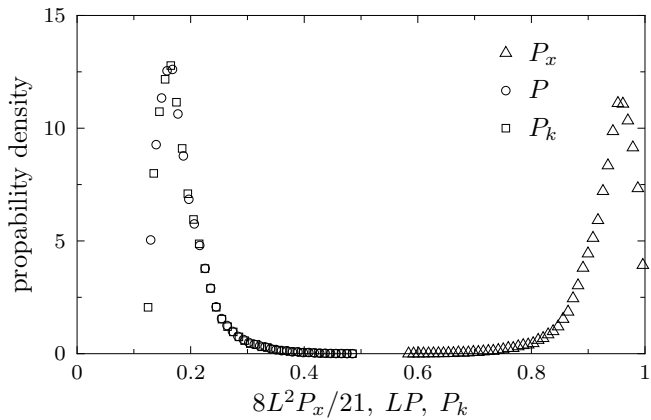}
\caption{Distributions of the IPRs in real space
(triangles), phase space (circles), and momentum space (squares) in
the zero disorder limit $W\to0$. A generic set of eightfold degenerate 
states in two dimensions with momentum components $\pm5\pi/24$ and 
$\pm3\pi/4$ is considered for a system of size $L=48$ where the overlap 
of the Husimi functions for stripes at $k_1$ and $k_2$ is negligible. The 
distributions are obtained by taking all eight degenerate states for 
10\,000 disorder realizations into account.}
\label{fig:iprw0}
\end{figure}

While for two degenerate states the limit $W\to 0$ 
leads to a universal value for the IPRs, this is no longer true in the
case of higher degeneracies, where the IPRs depend on the disorder 
realization, even in the limit $W\to 0$. This can already be seen
from the existence of different types of (real) wave functions. An 
optimal localization in momentum space can be obtained by pairing only
two plane waves with opposite momenta $\mathbf{k}$ and $-\mathbf{k}$,
leading to
\begin{equation}
\psi(\mathbf{x}) = (2/L)^{d/2}\cos(\mathbf{k}\cdot\mathbf{x})\,.
\label{eq:psi}
\end{equation}
A wide distribution in momentum space is achieved by a linear
combination of all energetically degenerate states. For a generic
state this yields
\begin{equation}
\phi(\mathbf{x}) = \frac{1}{(2^{d-2}d!L^d)^{1/2}}\sum_{P(\{k_i\})}
\sum_{\{\eta\}}\cos\left(\sum_{i=1}^d\eta_i k_i x_i\right)\,,
\label{eq:phi}
\end{equation}
where the first sum is to be taken over all permutations of the set of 
(different) momentum components $k_i$ while the second sum over the 
set of $\eta_i$ is to be taken over all combinations of factors $\pm1$ with
$\eta_1=+1$ kept fixed. Only for $d=1$ the states $\psi(x)$ and
$\phi(x)$ coincide, hinting again at the difference between the weak disorder
behavior in one dimension and the subtleties appearing in higher dimensions.

It follows that in two and higher 
dimensions a nontrivial distribution of IPRs already appears in the limit 
$W\to0$. Fig.~\ref{fig:iprw0} depicts such distributions for a set
of eightfold degenerate states in two dimensions with momentum components 
$\pm5\pi/24$ and $\pm3\pi/4$. The system size of $L=48$ ensures that the
overlap of the Husimi functions corresponding to the eight different 
momentum vectors is negligible. 

The two states (\ref{eq:psi}) and (\ref{eq:phi}) help to understand the
distribution for the momentum space IPR. On the one hand,
$\psi(\mathbf{x})$ yields the maximum IPR in momentum space for real wave 
functions, $P_k=1/2$. On the other hand, all plane waves might be equally 
weighted as in state $\phi(\mathbf{x})$, thus leading to an inverse
participation ratio of $1/8$ in momentum space. It turns out that the 
mixing of the plane waves due to a random potential is quite efficient 
thus making the first limit rather improbable.

For the case presented in Fig.~\ref{fig:iprw0}, the two momentum components 
$k_1$ and $k_2$ are well separated on the scale of the phase space 
resolution in momentum direction
$\sqrt{\pi/L}$. Consequently, the overlap of the resulting stripes in
phase space is negligible. Therefore, the distributions for the IPR in
momentum space and phase space coincide up to a scaling factor $L^{-d/2}$
which arises from the finite width of the Husimi function in the $d$
momentum directions.

In contrast to the behavior in momentum and
phase space, an equally weighted combination of all energetically degenerate 
plane waves leads to a maximum of the IPR in real space. For every pair of
different and non-zero momentum components $k_1$ and $k_2$, which are both a 
multiple of $2\pi/L$, the IPR for such a state becomes in an integral
approximation
\begin{equation}
P_x=\int_0^L\!d\mathbf{x}\,\phi(\mathbf{x})^4=\frac{21}{8L^2}\,.
\label{eq:iprxw02dmax}
\end{equation}
When only one momentum direction contributes, the opposite limit is
reached and the IPR in real space becomes
\begin{equation}
P_x=\int_0^L\!d\mathbf{x}\,\psi(\mathbf{x})^4=\frac{3}{2L^2}\,,
\label{eq:iprxw02dmin}
\end{equation}
which, up to a factor $1/L$, coincides with the result in one
dimension, cf.~Eq.~(\ref{eq:iprxw01d}). Since the equally weighted
states described by (\ref{eq:phi}) now lead to larger values of the
IPR, the result in real space is essentially a mirror image of the
IPRs in momentum and phase space (see Fig.~\ref{fig:iprw0}).

In three dimensions the IPRs for a generic situation of 48-fold degeneracy
may be obtained as well. While $L^{3/2}P$ and $P_k$ yield values between
$1/48$ and $1/2$, the IPR in real space assumes values between
$3/2L^3$ and $61/16L^3$. The latter values are obtained by a calculation
analogous to that underlying Eqs.~(\ref{eq:iprxw02dmax}) and 
(\ref{eq:iprxw02dmin}).

We emphasize once more that the states discussed above represent the
generic states. In addition, there exist states where some or all of
the momentum components are equal so that the number of degenerate
states is decreased. On the other hand, in certain cases a given total 
energy can be constructed by different sets of momentum components thus 
giving rise to an increase of the degeneracy. These special states are 
relevant for a detailed description of the complete distribution of IPRs 
for a given system size, which may exhibit a complex structure. 
However, in the limit of large system size, the generic states 
discussed above dominate the distributions.

From the perturbative investigation of the IPR in the limits of
very strong and very weak disorder, we can conclude
that real space properties will only dominate at weak disorder strength,
if a coupling is induced predominantly between plane waves close in
momentum. As long as the distance in momentum is below the momentum
uncertainty $\sqrt{\pi/L}$ in phase space, such a coupling will become
apparent only via the large scale real space structure appearing in the
Husimi function. Therefore, in this scenario which is characteristic
for one dimension, real space dominates at weak disorder strength and
the phase space IPR increases with increasing disorder. It is only the
value of $P(W=0)$ itself which is determined by momentum space properties.

In two and higher dimensions the picture changes drastically, because states 
close in energy are not necessarily close in momentum anymore. In this case,
the disorder potential may scatter plane waves into other momentum directions 
and thus induces a strong mixing in momentum space. In particular, states
of the type (\ref{eq:psi}), which yield large values for $P$, are affected 
by such processes. The mixing will thus lead to a decrease of the IPR in 
phase space as a consequence of the dominance of momentum space. In
contrast, the real space structure will appear on relatively small length 
scales which are typically not resolved by the Husimi function.
As we will see in the following section, this decrease of $P$ in $d\ge2$ 
implies the existence of a regime of intermediate disorder where phase
space is well covered by the Husimi function and which can be associated
with diffusive behavior. Furthermore, this scenario opens the possibility
of a delocalization-localization transition.

\section{System size dependence}
\label{sec:size}

\begin{figure}
\includegraphics[width=\columnwidth]{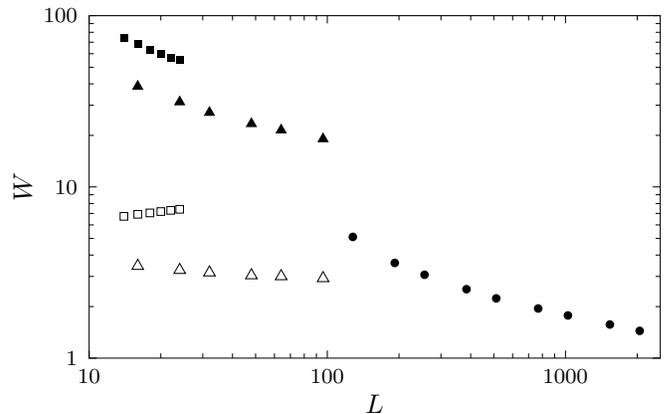}
\caption{Values of the disorder strength $W$ for the minima (open
symbols) and maxima (full symbols) of the inverse participation ratio
in phase space, as a function of the system size. The circles, triangles
and squares are for one, two and three dimensions, respectively.}
\label{fig:minmax}
\end{figure}

While the Anderson model in three dimensions exhibits a phase transition
from delocalized states at weak potential to localized states at strong
potential, such a phase transition is absent for the Anderson model in two
dimensions, where all states are localized in the thermodynamic limit.
Nevertheless, for fixed system size, both models show qualitatively
the same behavior of the IPR in phase space, calling for an analysis of the
size dependence of the IPR in order to check whether or not the strong
increase of the phase space IPR at the crossover between the diffusive and
the localized regime evolves towards an abrupt jump which would indicate a
phase transition in the limit $L\to\infty$.  

To this end we plot in Fig.~\ref{fig:minmax}, versus the system size $L$,
the disorder strengths $W$ at which for the Anderson model in one, two and
three dimensions maxima and minima of the phase space IPR occur. The
locations of minima and maxima are shown as open and full symbols,
respectively. While scaling laws cannot be extracted from the
data,\footnote{Due to numerical limitations, the range of available system
sizes in two and particularly in three dimensions is quite small. On the
one hand, the system size should be sufficiently large compared to the
width of the minimal uncertainty state (\ref{eq:koh}) in order to avoid
artifacts stemming from the periodic boundary conditions. On the other
hand, the numerical efforts required in the diagonalization of the
Hamiltonian~(\ref{eq:ham}) limit the system size from above.} one can
nevertheless clearly observe the direction of the shift of minima and
maxima as a function of the system size.  This reveals an important
difference between the cases of two and three dimensions. In three
dimensions, the position of the maximum moves to lower disorder values when
the system size is increased, while the position of the minimum shifts in
the opposite direction.  From this trend, one can expect that in the limit
$L\to\infty$, the positions of the maximum and the minimum converge towards
the same finite disorder value, with the emergence of a non-monotonic step
in the disorder-dependence of the phase space IPR as a clear signature of
the Anderson transition in phase space. 

\begin{figure}
\includegraphics[width=0.7\columnwidth]{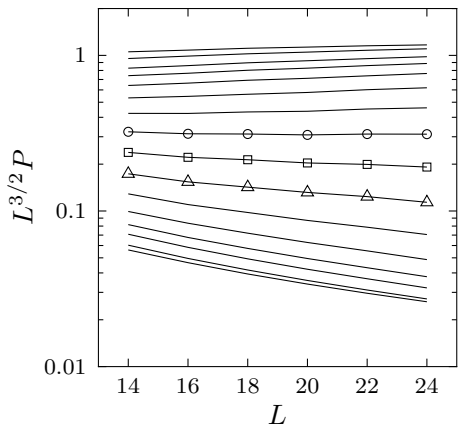}
\caption{Dependence of the phase space IPR on the system size for the
Anderson model in $d=3$ at fixed disorder strengths $W= 6.9$, $9.1$, $11.0$, 
$12.0$, $13.2$, $14.5$, $15.8$ (triangles), $17.4$ (squares), $19.1$ (circles), 
$20.9$, $22.9$, $25.1$, $27.5$, $30.2$, $36.3$, and $47.9$ from the lower 
to the upper curve.}
\label{fig:iprscale}
\end{figure}

In order to get an estimate of the critical disorder strength $W_c$ for the
Anderson model in $d=3$, we depict in Fig.~\ref{fig:iprscale} 
the change of the phase space IPR as a function of the system size for
fixed values of the disorder strength. For $W<W_c$, the phase space IPR
should decrease with increasing $L$ while for $W>W_c$ it increases. 
For $W=19.1$ (indicated by circles), $P$ will increase with the system
size. For $W=17.4$ (squares), $P$ decreases for the system sizes accessible
to us, but one may anticipate that the curve will rise for larger system sizes.
Such a behavior can be observed also for the two-dimensional Anderson
model. In contrast, for $W=15.8$ (triangles), one would expect that the
curve continues to fall even for larger system sizes. This implies a
critical value for the disorder strength between 15.8 and 17.4. While these
considerations are not necessarily stringent, the results presented in
Fig.~\ref{fig:iprscale} as well as in Figs.~\ref{fig:ipr123} and
\ref{fig:minmax} are perfectly consistent with the known value of 
$W_c\approx 16.5$ for the Anderson transition in the band center \cite{macki81}.

In contrast, according to the data depicted in Fig.~\ref{fig:minmax} for 
two dimensions, the positions of the minimum and of the maximum IPR both 
move towards lower disorder values when $L$ increases. It seems plausible 
that they both go to zero in the limit of infinite system size, consistent 
with an extension of the localized regime down to infinitesimal disorder 
strength and the absence of a phase transition in two dimensions. However, 
the fact that the overall behavior at finite system size in two and three 
dimensions is very similar, hints at the role of $d=2$ as a marginal 
dimension in the Anderson model.

A better insight into the behavior of the phase space IPR can be gained by
considering the position of the maxima, which shift to smaller disorder 
strength with increasing system size, independently of the
dimensionality. For a given system size $L$, the maximum phase space 
IPR appears in the localized regime at a certain disorder strength $W$. 
Now, for a localized state at this fixed disorder strength,
but at larger system size, the phase space IPR becomes independent of 
the spatial structure once the width of the minimal uncertainty 
state (\ref{eq:koh}) exceeds the localization length.
In this regime, the phase space IPR is dominated by momentum
space features. Since we know that $dP_k/dW<0$ 
we can conclude that $dP/dW<0$. Therefore, 
the maximum of $P(W)$ shifts to smaller disorder strength when the 
system size is increased. 
Furthermore, from this argument it follows that in one and two 
dimensions the maximum shifts to $W=0$ in the limit $L\to\infty$ 
if we infer that all states are localized in the thermodynamic limit.

In order to discuss the position of the minimum IPR,
we now turn to the ballistic regime at weak disorder. The coupling between
plane waves within first order perturbation theory depends on two contrary
effects. On the one hand, the number of plane waves into which scattering
may occur increases with the system size. On the other hand, since the
disorder potentials at different lattice sites are uncorrelated, the
individual coupling matrix elements decrease with the size of the system.
However, independently of the dimension the increased density of states
dominates and scattering becomes more effective as the system size
increases. This corresponds to a shrinking of the ballistic regime which
can be seen in Fig.~\ref{fig:ipr123} as a shift of the curves to smaller
$W$ with increasing $L$. This discussion, however, does not restrict the 
position of the minimum of $P$ as a function of system size since the
minimum always appears in the diffusive regime. Indeed, as 
Fig.~\ref{fig:minmax} shows, with increasing $L$,
the position of the minimum clearly shifts towards weaker disorder in 
two dimensions while it shifts to stronger disorder in three dimensions.

\section{Conclusions}
\label{sec:concl}

In this study of the phase space properties of the Anderson
model, we have demonstrated the potential immanent to this approach 
and, in particular, its advantages over approaches based purely 
on real or momentum space properties. In contrast to the latter ones, 
the phase space approach allows to treat real and momentum space on the same
footing. The well-studied Anderson model has allowed us to establish
an interpretation of the phase space IPR which will be useful in
cases where no independent information is available. 

We found that the crossover between the diffusive and the localized
regimes is accompanied by an increase of the phase space IPR which,
in three dimensions, evolves to a sharp step in the thermodynamic limit. 
This is a signature of the Anderson metal-insulator transition. 

The jump of the phase space IPR at the Anderson transition implies a
dramatic reorganization of the Husimi distribution from a large spread over
phase space to localization not only in real space but also in phase space.
This scenario is not only relevant for the $d=3$ Anderson model, but
corresponds to the very similar one that was recently found for the
Aubry-Andr{\'e} model \cite{ingol02}. It is advantageous to exploit this
similarity. The one-dimensional Aubry-Andr{\'e} model allows for a direct
visualization of the changes in the Husimi function at the metal-insulator
transition.  Furthermore, in numerical treatments of the Aubry-Andr{\'e}
model, the system size may be varied by more than two orders of magnitude,
thus allowing for a much more detailed study of the
phase transition \cite{aulbachxx}.

Moreover, by putting together the insights gained from phase space into the
Aubry-Andr{\'e} model and the Anderson model at different dimensions,
it becomes clear that the dimensionality of the model is not the most
important parameter for the occurrence of a phase transition.
Instead, we could identify the disorder-induced coupling of plane waves
having distant momenta as the relevant mechanism for the
occurrence of a phase transition. 

It will be interesting to apply these phase space concepts to interacting
systems where the possibility to characterize individual many-particle
states is expected to be of great value. Work along these lines is in
progress.

\begin{acknowledgments}
This work was supported by the Son\-der\-for\-schungs\-be\-reich 484 of the
Deutsche Forschungsgemeinschaft. D.W. thanks the European Union for
financial support within the RTN program. The numerical calculations were 
carried out partly at the Leibniz-Rechenzentrum M{\"u}nchen.
\end{acknowledgments}

\end{document}